\documentclass[12pt]{article}
\usepackage{fancyhdr}
\usepackage{latexsym} 
\usepackage{amssymb}
\usepackage{graphicx}
\usepackage{epsf}
\usepackage{amscd}
\usepackage{amsmath}

\makeatletter
\@addtoreset{equation}{section}

\renewcommand{\thefootnote}{\fnsymbol{footnote}}
\makeatother

\begin{document}
\thispagestyle{empty}

\begin{flushright}
TIT/HEP--537 \\
{\tt hep-th/0505136} \\
May, 2005 \\
\end{flushright}
\vspace{3mm}

\begin{center}
{\Large \bf 
Moduli Space of BPS Walls \\
in Supersymmetric Gauge Theories 
} 
\\[12mm]
\vspace{5mm}

\normalsize
  {\large \bf 
Norisuke~Sakai$^1$}
\footnote{\it  e-mail address: 
nsakai@th.phys.titech.ac.jp
} 
~and~~  {\large \bf 
Yisong~Yang$^2$}
\footnote{\it  e-mail address: yyang@math.poly.edu
} 

\vskip 1.5em

{ \it $^1$Department of Physics, Tokyo Institute of 
Technology \\
Tokyo 152-8551, JAPAN  
 }
\\
~and~~  
\\
{\it $^2$Department of Mathematics, Polytechnic University \\
Brooklyn, New York 11201, U.S.A. 
 }
\vspace{15mm}
{\bf Abstract}\\[5mm]
{\parbox{13cm}{\hspace{5mm}
Existence and uniqueness of the solution are 
proved for the `master equation' derived from 
the BPS equation for 
the vector multiplet scalar in the $U(1)$ 
gauge theory with $N_{\rm F}$ charged matter 
hypermultiplets with eight supercharges. 
This proof establishes that the solutions of the BPS 
equations are completely characterized by 
the moduli 
matrices divided by the $V$-equivalence relation 
for the gauge theory at finite gauge couplings. 
Therefore the moduli space at finite gauge 
couplings is topologically the same manifold as that at 
infinite gauge coupling, where the gauged linear sigma 
model reduces to a nonlinear sigma model. 
The proof is extended to 
the $U(N_{\rm C})$ gauge theory with 
$N_{\rm F}$ hypermultiplets in the fundamental 
representation, provided 
the moduli matrix of the domain 
wall solution is $U(1)$-factorizable. 
 Thus the dimension of 
the moduli space of $U(N_{\rm C})$ gauge theory is 
bounded from below by the dimension of the 
$U(1)$-factorizable part of the moduli space.  
We also obtain sharp estimates of the asymptotic exponential 
decay which depend on both the gauge coupling 
and the hypermultiplet mass differences.
}}
\end{center}
\vfill
\newpage
\setcounter{page}{1}
\setcounter{footnote}{0}
\renewcommand{\thefootnote}{\arabic{footnote}}

\section{Introduction}\label{INTRO}

Solitons have been important 
in understanding nonperturbative effects in field 
theories \cite{SW}. 
They are also useful to construct models of the 
brane-world scenario \cite{HoravaWitten}--\cite{RandallSundrum}. 
The simplest of these solitons is the domain wall 
separating two domains of discretely different vacua. 
The supersymmetric theories are useful to  obtain  
realistic unified theories beyond the standard model\cite{DGSW}. 
If a field configuration preserves a part of supersymmetry, 
it satisfies the field equation automatically \cite{WittenOlive}. 
Such a configuration is called the Bogomol'nyi-Prasad-Sommerfield 
(BPS) configuration \cite{BPS}. 
The BPS domain walls have been much studied in supersymmetric 
field theories with four 
supercharges \cite{Abraham:1992vb}, \cite{DvaliShifman}, 
and with eight supercharges\cite{multi}-\cite{SakaiTong}. 
These soliton solutions often contain parameters, which are 
called moduli. 
If we promote the moduli  parameters as fields on 
the world volume of the soliton, they give massless fields  
on the world volume of the soliton \cite{Manton:1981mp}. 
The metric on the moduli space gives a 
(nonlinear) kinetic term of the Lagrangian of the 
low-energy effective field theory. 
Therefore the determination of the moduli space is of vital 
importance to  understand  the dynamics of the solitons. 

One of the most interesting classes of models possessing 
domain walls is the gauge theories with eight 
supercharges\cite{Tong}-\cite{global}. 
To allow domain walls, we need discrete vacua. 
For that purpose, we introduce Fayet-Iliopoulos (FI) terms 
for a $U(1)$ factor gauge group. 
As a natural gauge group with the $U(1)$ factor, we choose 
$U(N_{\rm C})$ gauge theory. 
For simplicity, we take the matter hypermultiplets in the 
fundamental representation of $U(N_{\rm C})$. 
To obtain more than one supersymmetric vacua, we require 
that the number of  flavors of hypermultiplets $N_{\rm F}$ 
be larger than the number of colors  $N_{\rm C}$ 
\begin{equation}
N_{\rm F} > N_{\rm C}
\label{eq:color-flavor}
\end{equation}
With massless hypermultiplets\footnote{
A common mass can be absorbed into the shift of vector 
multiplet scalar and is not relevant.
}, the vacuum manifold is a hyper-K\"ahler manifold, 
the cotangent bundle over the complex Grassmann manifold 
$T^*G_{N_{\rm F}, N_{\rm C}}$. 
It reduces to $T^*CP^{\rm N_{\rm F}-1}$ manifold in the 
case of the $U(1)$ gauge theory ($N_{\rm C}=1$). 
If the nondegenerate hypermultiplet masses are turned on, 
a potential term is induced and most of the vacua are 
lifted, allowing wall solutions. 
In the resulting vacua of the massive $U(N_{\rm C})$ gauge 
theories, each color component of the hypermultiplets 
chooses to have a particular flavor 
(color-flavor-locking) \cite{ANS}, \cite{shif2}. 
Furthermore, a systematic construction of BPS wall solutions 
has been established \cite{INOS1}, \cite{INOS2}.  

If we take the limit of strong gauge coupling 
$g^2\rightarrow \infty$ for the $U(1)$ gauge theory, 
the vector multiplet can be eliminated to give 
constraints on the hypermultiplet field space, resulting 
in a supersymmetric massive nonlinear sigma model with 
the $T^*CP^{N_{\rm F}-1}$ target space \cite{LR}, 
\cite{HKLR}. 
Equations for preserving half of supersymmetry 
are called the $1/2$ BPS equations. 
As boundary conditions for the BPS 
equations, the field configurations are 
required to approach one of the discrete vacua. 
If the vacua at $y=-\infty$ and $+\infty$ happen to be 
different, the solution represents (multi-) walls. 
If the vacua at $y=-\infty$ and $+\infty$ happen to be 
identical, the solution represents one of vacua. 
Therefore these BPS equations admit vacua (full 
supersymmetry conserved) besides (multi-) walls as 
solutions. 
The physical relation between these different topological 
sectors are as follows. 
If we let the position of one of the walls to go to 
infinity, we obtain a topological sector with one less 
wall. 
Therefore the topological sectors with $n-1$ walls 
appear as boundaries of the moduli space of a topological 
sector with $n$ walls. 
Continuing this process, we eventually arrive at 
topological sectors with no walls, namely vacua. 
In this way, we naturally obtain a compactification of 
moduli space of various topological sectors of multi-walls. 
The resulting manifold of all the solutions of the $1/2$ 
BPS equations is topologically  $CP^{N_{\rm F}-1}$ in our 
case of the $T^*CP^{N_{\rm F}-1}$ nonlinear sigma model 
\cite{multi}, \cite{ANNS}. 
The massive $U(N_{\rm C})$ gauge theory 
reduces to the massive nonlinear sigma model with the 
$T^*G_{N_{\rm F}, N_{\rm C}}$ target space and with 
a potential term in the limit of strong gauge coupling 
\cite{ANS}. 
Similarly to the $U(1)$ case, the space of all solutions 
of the $1/2$ BPS equations for the massive 
$T^*G_{N_{\rm F}, N_{\rm C}}$ nonlinear sigma model 
with the vacuum boundary condition at infinity 
is found to be the complex Grassmann manifold 
$G_{N_{\rm F}, N_{\rm C}}$, which is 
the special Lagrangian submanifold of the 
target space of the nonlinear sigma model 
\cite{INOS1}, \cite{INOS2}. 
More generally, it has been found that there are 
nonlinear sigma models with a target space 
having several special Lagrangian submanifolds whose union 
gives the space of all solutions of the $1/2$ BPS 
equations \cite{global}.  

A number of exact solutions for $U(1)$ gauge theories 
have been obtained 
for particular discrete finite values of gauge coupling 
\cite{KakimotoSakai}, \cite{IOS1}, \cite{INOS2}. 
For generic finite gauge couplings of $U(1)$ as well as 
$U(N_{\rm C})$ gauge theories, only the general behavior of 
domain walls have been studied qualitatively 
\cite{shif}, \cite{shif2}, \cite{SakaiTong}. 
The systematic construction of BPS walls first solves the 
hypermultiplet BPS equation and yields 
the moduli matrix $H_0$ as integration constants. 
After taking account of  an  equivalence relation called 
the $V$-equivalence relation, 
the independent variables in the moduli matrix $H_0$ 
constitute the complex Grassmann manifold 
$G_{N_{\rm F}, N_{\rm C}}$. 
The remaining BPS equation for the vector multiplet scalar 
can be rewritten  into a ``master equation'', which is a  
nonlinear ordinary differential equation 
for a gauge invariant quantity $\Omega$. 
In the limit $g^2 \rightarrow \infty$, 
this  master equation  can be solved algebraically 
without introducing additional moduli, once the moduli 
matrix $H_0$ is given. 
It has been conjectured that there exists a unique solution 
of the  master equation  even at finite gauge coupling, 
once the moduli matrix $H_0$ is given. 
Based on this conjecture, it has been pointed out 
that the moduli space of the BPS equations is given by 
the moduli matrix $H_0$ divided by the $V$-equivalence 
relation even at finite gauge couplings 
\cite{INOS1}, \cite{INOS2}. 
Up to now, the best supporting evidence for this proposal 
is given by the index theorem, although the evidence is only 
indirect. 
The index theorems are proven for 
$U(1)$ gauge theories \cite{keith}, \cite{global}, 
and for $U(N_{\rm C})$ gauge 
theories\cite{SakaiTong}, respectively. 
They state that the complex Grassmann manifold 
contains necessary and sufficient number of moduli parameters. 
However, it is much more desirable to demonstrate the 
existence and uniqueness of the solution of the 
 master equation. 

The purpose of this paper is to study the 
 master equation for  
the gauge invariant quantity $\Omega$. 
We present a proof of the existence and 
uniqueness of the solution for the $U(1)$ gauge theories, 
and extend the proof to a class of the 
moduli matrix $H_0$ in the $U(N_{\rm C})$ gauge theories. 
Our proof for the $U(1)$ case finally establishes that 
solutions of the $1/2$ BPS equations 
are completely characterized by 
the moduli matrices divided by the 
$V$-equivalence relation. 
This moduli space is topologically the same as 
the moduli space of the BPS equations in the 
nonlinear sigma 
model with $T^*CP^{N_{\rm F}-1}$ target space, which is 
obtained in the limit of $g^2 \rightarrow \infty$. 
Of course the metric on the moduli space at finite gauge 
coupling is expected to be 
deformed from that of nonlinear sigma model (infinite gauge 
coupling). 
We also obtain  estimates of the asymptotic 
exponential decay which depend on both the hypermultiplet 
mass differences and the gauge coupling squared 
multiplied by the FI-parameter. 
These estimates agree   with our previous result based on an iterative 
approximation scheme \cite{IOS1}. 
For the non-Abelian $U(N_{\rm C})$ gauge theories, we show 
that the proof can be applied to the part of the moduli space 
which is described by the $U(1)$ factorizable moduli matrix 
$H_0$. 
This result implies a lower bound of the dimensions of the 
moduli space for the $U(N_{\rm C})$ gauge theories. 
We will leave for future publication 
to extend the proof of the existence and uniqueness 
of the  master equation for the gauge invariant 
$\Omega$ to entire 
moduli space of the BPS walls in the $U(N_{\rm C})$ gauge 
theories. 

 Interestingly, our one-dimensional nonlinear equation and 
its variational structure resemble in many ways the 
two-dimensional Abelian BPS vortex equation which allows us 
to extend the method developed in Jaffe and Taubes 
\cite{JT} to our problem here. There are two major 
technical differences/difficulties, though, that need to 
be overcome. 
The first one is that in one dimension, the ranges of the 
exponents in the Gagliardo--Nirenberg inequality cannot 
render as strong an estimate as in two dimensions (a 
relevant lower bound takes the weaker, sublinear, form, 
$\|v\|_2^{2/3}$ instead of the usual stronger, linear, 
form, $\|v\|_2$. See (\ref{eu13})). 
The second one is that, unlike in the Abelian vortex 
situation in which the vacuum state is uniquely 
characterized by the asymptotic amplitude of the Higgs 
field, our domain wall solution needs to interpolate two 
different vacua at the two infinities of the real line. 
Hence, the behavior of the solution in a local region 
is less uniform and the decay rates near the two infinities 
are also necessarily different.

In section \ref{sc:u1}, 
BPS equations and their systematic solutions are introduced. 
We also describe implication of the existence and uniqueness 
of the  master  equation for the gauge invariant. 
In section \ref{sc:uN}, 
we extend our analysis to the $U(1)$ factorizable case of 
the $U(N_{\rm C})$ gauge theories, giving a lower bound 
for the dimension of the BPS wall moduli space. 
In section \ref{sc:existence}, we present 
an analytic proof of the existence and uniqueness of the 
 master  equation for the gauge invariant, and 
give  estimates  
of the asymptotic behavior of the solution.

\section{
Moduli Space of BPS Equations in $U(1)$ 
Theories
}
\label{sc:u1}

Let us take a supersymmetric $U(1)$ gauge theory 
with eight supercharges in one time and four spatial 
dimensions\footnote{
Discrete vacua are required for wall solutions and are 
possible by mass terms for hypermultiplets which 
are available only in spacetime dimensions equal to or 
less than five. 
}. 
The $U(1)$ vector multiplet contains 
gauge field $W_M$, gaugino $\lambda^i$, 
a real neutral scalar 
field $\Sigma$, and $SU(2)_R$ triplet of 
real auxiliary fields $Y^a$, where 
$M,N=0,1,\cdots,4$ denote space-time indices, and 
 $i=1, 2$ and $a=1,2,3$ denote $SU(2)_R$ doublet 
and triplet indices,  respectively. 
The hypermultiplet contains two 
complex scalar fields $H^{iA}$, hyperino 
$\psi^A$ and 
complex auxiliary fields $F_i^A$, where 
$A=1, \cdots, N_{\rm F}$ 
stand for flavors. 
For simplicity, we assume that these $N_{\rm F}$ 
hypermultiplets have the same $U(1)$ charge, say, 
unit charge. 
Denoting the gauge coupling $g$, 
the mass of the $A$-th 
hypermultiplet $m_A$, and the FI parameters $\zeta^a$, 
the bosonic part of the Lagrangian 
reads 
\cite{Tong}--\cite{KakimotoSakai} 
\begin{eqnarray}
\mathcal{L}_{\rm boson}
&\!\!\!=&\!\!\!
-\frac{1}{4g^2}(F_{MN}(W))^2
+\frac{1}{2g^2}({\partial}_M\Sigma )^2 
+({\cal D}_MH)^\dagger_{iA}({\cal D}^MH^{iA})
-H^\dagger_{iA}(\Sigma-m_A)^2H^{iA} 
\nonumber\\
&\!\!\!&\!\!\!
+\frac{1}{2g^2}(Y^a)^2 - \zeta^aY^a 
+ H^\dagger_{iA}(\sigma^aY^a)^i{}_jH^{jA} 
+F^{\dagger i}_AF_i^A ,
\label{5DW-3.1}
\end{eqnarray}
where a sum over repeated indices is understood, 
$F_{MN}(W)=\partial_M W_N -\partial_N W_M$, 
covariant derivative is defined as 
${\cal D}_M =\partial_M + iW_M$, 
and our metric is $\eta_{MN}=(+1, -1, \cdots, -1)$. 
We assume that the hypermultiplet masses are nondegenerate 
and are ordered as 
\begin{equation}
m_{1}>m_2> \cdots >m_{N_{\rm F}}. 
\label{5DW-3.6}
\end{equation}
The auxiliary fields are given by their equations of motion: 
$F_i^A=0 $ and 
\begin{eqnarray}
Y^a&\!\!\!=&\!\!\!
g^2[ \zeta^a-H^\dagger_{iA}(\sigma^a)^i{}_jH^{jA}].
\label{eq:EOMauxilary}
\end{eqnarray}
By making an $SU(2)_R$ transformation, we can choose 
the FI parameters to the third direction 
\begin{equation}
\zeta^a=(0,0,\zeta), \quad \zeta >0.
\label{eq:FIterm}
\end{equation}
In this choice, we find $N_{\rm F}$ discrete SUSY vacua 
($A=1, \cdots, N_{\rm F}$) as 
\begin{eqnarray}
\Sigma = m_A,&& |H^{1A}|^2=\zeta,\quad H^{2A}=0 , \nonumber\\
&& H^{1B}=0 ,\quad\quad H^{2B}=0, \quad\quad (B\not = A). 
\label{5DW-3.11}
\end{eqnarray}

We assume the configuration to depend only on a single 
coordinate, which we denote as $y \equiv x^4$, 
and assume the 
four-dimensional Lorentz invariance in the world volume 
coordinates $x^\mu=(x^0, \cdots, x^3)$. 
Let us examine the supersymmetry transformations of fermions: 
the gaugino $\lambda^i$ and hyperino $\psi^A$ 
transform as\footnote{ 
Our gamma matrices are $4\times 4$ matrices and are defined as : 
$\{ \gamma^M,\gamma^N\}=2\eta^{MN }$, 
$\gamma^{MN}\equiv \frac{1}{2}[\gamma^M,\gamma^N]=\gamma^{[M}\gamma^{N]}$, 
$\gamma^5\equiv i\gamma^0\gamma^1\gamma^2\gamma^3=
-i \gamma^4$. 
}
\begin{eqnarray}
\delta_\varepsilon \lambda^i 
&\!\!\!=&\!\!\! \Bigl(\frac{1}{2}\gamma^{MN}F_{MN}(W)
+\gamma^M{\partial}_M\Sigma \Bigr)\varepsilon^i
+i\Bigl(Y^a\sigma^a\Bigr)^i{}_j\varepsilon^j, 
\label{eq:gauginoSUSY}
\\
 \delta_\varepsilon \psi^A 
&\!\!\!=&\!\!\! -i\sqrt{2} \Bigl[ \gamma^M{\cal D}_MH^{iA} 
+i(\Sigma-m_A) H^{iA} \Bigl] \epsilon_{ij}\varepsilon^j
+\sqrt{2}F_i^A\varepsilon^i .
\label{eq:SUSYtrans}
\end{eqnarray}
We require the following half of supersymmetry 
to be preserved 
\begin{equation}
P_+ \varepsilon^1 =0, \quad P_-\varepsilon^2 =0 ,
\label{5DW-3.13}
\end{equation}
where $P_{\pm}\equiv (1 \pm \gamma_5)/2$ are the chiral 
projection operators. 
Then we obtain the ${1 \over 2}$ BPS 
equations 
\begin{equation}
{\cal D}_y H^{1A} =(m_A-\Sigma )H^{1A}, 
\quad 
{\cal D}_y H^{2A} =(-m_A+\Sigma )H^{2A}, 
\quad A=1, \cdots, N. 
\label{5DW-3.4}
\end{equation}
\begin{equation}
0=Y^1+iY^2=-2g^2 H^{2A} (H^{1A })^*
, 
\label{eq:BPS2}
\end{equation}
\begin{equation}
\partial_y \Sigma 
=
Y^3=g^2 
\left(\zeta-H_{1A}^\dagger H^{1A}+H_{2A}^\dagger H^{2A} \right),
\label{5DW-3.3}
\end{equation}

We wish to obtain solutions of these BPS equations which 
interpolate two different vacua in Eq.(\ref{5DW-3.11}). 
The boundary condition of these two vacua at $y=\pm\infty$ 
specifies the topological sector. 
By letting the outer-most wall to infinity, one can obtain 
 topological sectors with one less wall. 
Therefore we are interested in the maximal topological sector 
which allows the maximal number of walls and possesses the 
maximal number of moduli parameters . 
The boundary 
conditions for the maximal topological sector 
are given by 
\begin{eqnarray}
\Sigma(-\infty)&\!\!\!
=&\!\!\!m_{N_{\rm F}},\quad \Sigma(\infty)=m_{1}, 
\label{eq:vector-bound-cond}
\\
H^{1A}(-\infty) &\!\!\!
=&\!\!\! \sqrt{\zeta}\delta_{N_{\rm F}}^A
,\quad H^{1A}(\infty)=\sqrt{\zeta}\delta_1^A, \nonumber\\
H^{2A}(-\infty)&\!\!\!
=&\!\!\!0,\quad H^{2A}(\infty)=0 .
\label{eq:hyper-bound-cond}
\end{eqnarray}

Let us define \cite{Tong}, 
\cite{IOS1}, \cite{INOS1} an $GL(1, {\bf C})$ 
group element $S(y)$ that expresses the vector multiplet 
scalar and gauge field as a pure gauge 
\begin{equation}
{1 \over S(y)} {d S(y) \over d y} = \Sigma(y) + i W_4(y) . 
\label{tong-15}
\end{equation}
The BPS equation for hypermultiplets (\ref{5DW-3.4}) 
can be solved in terms 
of this single complex function $S(y)$ as 
\begin{eqnarray}
H^{1A}(y) = S^{-1}(y)H_0^A  e^{m_A y}, 
\label{tong-13}
\end{eqnarray}
where complex integration constants $H_0^A$ can be assembled 
into a constant complex vector 
$H_0=(H_0^1, \cdots, H_0^{N_{\rm F}})$. 
This  $N_{\rm F}$ component vector is nothing but the 
$N_{\rm C}=1$ case of a 
$N_{\rm C}\times N_{\rm F}$ complex constant matrix 
called the moduli matrix \cite{INOS1}. 
Since Eq.(\ref{tong-15}) defines the function $S(y)$ only up to 
a complex multiplicative constant, equivalent descriptions of 
physical fields ($H^i$ and $\Sigma$) result from 
two complex vectors $H_0$ which 
are different by multiplication of a non-vanishing 
complex constant $V$ : 
\begin{eqnarray}
H_0\rightarrow V H_0, \qquad S\rightarrow V S, 
\label{eq:world-v-sym}
\end{eqnarray}
which is called the $V$-equivalence relation \cite{INOS1}. 
Therefore the moduli matrix $H_0$ in this case is 
topologically $CP^{N_{\rm F}-1}$. 
We define a $U(1)$ local gauge invariant function $\Omega(y)$ as 
\begin{eqnarray}
\Omega(y) = S(y)S^{*}(y). 
\label{eq:omega}
\end{eqnarray}
Using the hypermultiplet solution (\ref{tong-13}), 
the remaining BPS equation for the vector multiplet scalar 
can be rewritten in terms of $\Omega$ 
\begin{eqnarray}
{d \over dy}\left({1 \over \Omega(y)}{d\Omega(y) \over dy}\right) 
= g^2 \zeta 
\left( 1-{\Omega_0(y) \over \Omega(y)} \right), 
\label{eq:omega-master-eq-u1}
\end{eqnarray}
\begin{eqnarray}
\Omega_0(y)\equiv {1 \over \zeta}
\sum_{A=1}^{N_{\rm C}} H_0^A e^{2m_Ay} (H_0^A)^{*} 
\equiv e^{2W(y)}. 
\label{eq:omega0}
\end{eqnarray}
The boundary conditions in Eqs.(\ref{eq:vector-bound-cond}) 
and (\ref{eq:hyper-bound-cond}) now become the following 
boundary conditions for the master equation 
\begin{eqnarray}
\Omega(y) &\!\!\!\rightarrow&\!\!\! \Omega_0(y) \rightarrow 
(e^{2m_1y}|H_0^1|^2/\zeta, 0, \cdots, 0), 
\quad y\rightarrow \infty \nonumber \\
\Omega(y) &\!\!\!\rightarrow&\!\!\! \Omega_0(y) \rightarrow 
( 0, \cdots, 0, {e}^{2m_{N_{\rm F}}y}|H_0^{N_{\rm F}}|^2/\zeta), 
\quad y\rightarrow -\infty . 
\label{eq:boundary-cond-u1}
\end{eqnarray}
The non-vanishing left-most element of the moduli matrix 
$H_0$ specifies the boundary condition at $y=+\infty$, 
and the non-vanishing right-most element specifies the 
boundary condition at $y=-\infty$. 
We can see that the boundary condition is encoded in 
the choice of the moduli matrix $H_0$. 
The moduli matrix $H_0$ with more than one 
non-vanishing elements gives a (multi-) wall solutions, 
whereas the moduli matrix with a single non-vanishing 
element gives a vacuum solution. 
Therefore the space of all possible moduli matrix $H_0$ 
divided by the $V$-equivalence relation automatically 
gives all possible solutions of the BPS equation including 
the vacuum solutions and multi-wall solutions. 

 Using $W(y)$ in Eq.(\ref{eq:omega0}), we finally obtain 
{\it the master equation for the $U(1)$ 
gauge theory} \cite{IOS1}, \cite{To1} 
\footnote{We have changed our notation Re$\psi \rightarrow \psi$ 
from Ref.\cite{IOS1}. 
} 
\begin{eqnarray}
 \frac{1}{\zeta g^2}{d^2 \psi \over dy^2}
&\!\!\!=&\!\!\!
1-e^{-2 \psi(y) +2W(y)}, 
\quad 
\psi(y)\equiv {1 \over 2}\log \Omega(y). 
\label{eq:master-u1} 
\end{eqnarray}
The boundary conditions (\ref{eq:boundary-cond-u1}) 
now become the following 
boundary conditions for the master equation 
\footnote{We observe that one out of two constants 
$\log (|H_0^{1}|/\sqrt{\zeta})$ and 
$\log (|H_0^{N_{\rm F}}|/\sqrt{\zeta})$ can be 
absorbed into a shift of $\psi$. 
This freedom corresponds to the $V$-equivalence relation 
implying that only one out of these two 
constants in the boundary conditions 
is the genuine moduli parameter.} 
\begin{eqnarray}
\label{eq:bpund-cond-psi1}
\psi &\!\!\!\rightarrow&\!\!\! 
m_1 y+\log \left({|H_0^1|\over \sqrt{\zeta}}\right), 
\quad y \rightarrow +\infty, 
\\
\label{eq:bpund-cond-psi2}
\psi &\!\!\!\rightarrow&\!\!\! 
m_{N_{\rm F}} y+\log 
\left({|H_0^{N_{\rm F}}|\over \sqrt{\zeta}}\right), 
\quad y \rightarrow -\infty. 
\label{tong-18}
\end{eqnarray}

In Ref.\cite{INOS1}, it has been conjectured that there exists 
a unique solution of the  master equation 
(\ref{eq:master-u1}) given the 
boundary conditions (\ref{eq:bpund-cond-psi1}) and 
(\ref{eq:bpund-cond-psi2}). 
We shall prove the conjecture in section \ref{sc:existence}. 
With this proved, we can now state that the moduli space of the 
domain walls in the $U(1)$ gauge theory with $N_{\rm F}$ flavors 
is given by $CP^{N_{\rm F}-1}$ as described by the moduli matrix 
$H_0$ furnished with the equivalence relation (\ref{eq:world-v-sym}).

\section{
$U(1)$ Factorizable case of $U(N_{\rm C})$ 
Gauge Theories
}
\label{sc:uN}

Let us now turn our attention to a supersymmetric 
$U(N_{\rm C})$ gauge theory with $N_{\rm F} (> N_{\rm C})$ 
flavors of hypermultiplets in the fundamental representation. 
We denote gauge fields $W_M$ and real scalar fields $\Sigma$ 
as $N_{\rm C}\times N_{\rm C}$ matrix whose basis is normalized 
as  
\begin{eqnarray}
{\rm Tr}(T_I T_J) 
= {1 \over 2}\delta_{IJ}, \ \ \ [ T_I , T_J]
=i f_{IJ}{}^{K} T_K,
\ \ \  
T_0\equiv \sqrt{1 \over 2N_{\rm C}}
\label{norm-comm}
\end{eqnarray}
We consider a five-dimensional spacetime 
and nondegenerate masses for hypermultiplets as ordered 
in Eq.(\ref{5DW-3.6}). 
Hypermultiplets are denoted as $N_{\rm C}\times N_{\rm F}$ 
matrices $H^i$. 
The bosonic part of the Lagrangian of the supersymmetric 
$U(N_{\rm C})$ gauge theory reads  
\begin{eqnarray}
{\cal L}|_{\rm bosonic}&=&
 {\rm Tr}\biggl[-{1\over 2g^2}F_{MN}(W)F^{MN}(W)+
{1\over g^2}({\cal D}_M\Sigma )^2+{1\over g^2}(Y^a)^2
\nonumber \\
&&\qquad\qquad   +{\cal D}^MH^i{\cal D}_MH^{i\dagger }
-(\Sigma H^i-H^iM)(\Sigma H^i-H^iM)^\dagger +F^iF^{i\dagger}
\nonumber\\
&&\qquad\qquad 
{}-Y^a(c_a-(\sigma _a)^i{}_jH^jH^{i\dagger })\biggr]
\end{eqnarray}
with the Fayet-Iliopoulos parameter $c_a=(0,0,c)$ with $c>0$ 
and the 
hypermultiplet mass matrix 
$M={\rm diag}(m_1,\,\cdots, m_{N_{\rm F}})$.
Covariant derivatives are defined as 
${\cal D}_M \Sigma = \partial_M \Sigma + i[ W_M , \Sigma ]$, 
${\cal D}_M H^{irA}
=(\partial_M \delta_s^r + i(W_M)^r{}_s)H^{isA}$, 
and the gauge field strength is 
$F_{MN}(W)=-i[{\cal D}_M , {\cal D}_N]
=\partial_M W_N -\partial_N W_M + i[W_M, W_N]$. 

Considering the supersymmetry transformations in 
Eqs.(\ref{eq:gauginoSUSY}) 
and (\ref{eq:SUSYtrans}), and requiring the half of 
supersymmetry in Eq.(\ref{5DW-3.13}) 
to be preserved, 
we obtain the $1/2$ BPS equations for walls with profile in 
$y$ 
\begin{eqnarray}
{\cal D}_y H^1 &=& -\Sigma H^1 + H^1 M,
\label{eq:BPSeq-H1}\\ 
{\cal D}_y H^2 &=& \Sigma H^2 -H^2 M, 
\label{eq:BPSeq-H2}\\
{\cal D}_y \Sigma &=& Y^3 
={g^2\over 2}\left(c{\bf 1}_{N_{\rm C}}-H^1H^1{}^\dagger 
+H^2H^2{}^\dagger \right),
\label{BPSeq-Sigma}\\
0&=&Y^1+iY^2= - g^2 H^2H^1{}^\dagger . 
\label{BPSeq-Y12}
\end{eqnarray}
The supersymmetric vacua are characterized by the vanishing 
of the right-hand side of all of these BPS equations. 
The supersymmetric vacua have been found to be 
color-flavor-locked and discrete \cite{ANS}. 
For each color component $r$, only one 
flavor $A_r$ of hypermultiplet $H^{irA}, A=A_r$ 
should take a non-vanishing value, 
\begin{eqnarray}
 H^{1rA}=\sqrt{c}\,\delta ^{A_r}{}_A,\quad H^{2rA}=0,
 \label{eq:hyper-vacuum}
\end{eqnarray}
and the corresponding color component of the vector 
multiplet scalar $\Sigma$ should be equal to the mass of 
that flavor of the hypermultiplets 
\begin{eqnarray}
\Sigma ={\rm diag.}(m_{A_1},\,m_{A_2},\,\cdots,\,
m_{A_{N_{\rm C}}}).
 \label{eq:vector-vacuum}
\end{eqnarray}
We denote the above SUSY vacuum as 
\begin{eqnarray}
 \langle A_1\,A_2\,\cdots\,A_{N_{\rm C}}\rangle .
\end{eqnarray}
Consequently topological sector is labeled by two vacua : 
\begin{equation}
\langle A_1, \cdots, A_{N_{\rm C}} \rangle \leftarrow 
\langle B_1, \cdots, B_{N_{\rm C}} \rangle, 
\end{equation}
where the first vacuum 
is at $y=\infty$ and the second at $y=-\infty$. 
Ideally we wish to solve the BPS equations for each 
topological sector and we wish to obtain all possible 
solutions. 

Let us consider the maximal topological 
sector, which is specified in the $U(N_{\rm C})$ gauge 
theory as : 
\begin{equation}
\langle 1, \cdots, {N_{\rm C}} \rangle \leftarrow 
\langle N_{\rm F}-N_{\rm C}-1, \cdots, N_{\rm F} \rangle. 
\end{equation}
In this maximal topological sector, the boundary conditions 
for the hypermultiplet scalar $H^1(y)$ is given by 
\begin{eqnarray}
 H^1(y)\rightarrow 
\sqrt{c}\left(
\begin{array}{ccccccc}
1     &0     &\cdots &0     &0     &\cdots&0      \\
0     &1     &\cdots &0     &0     &\cdots&0      \\
\vdots&\vdots&       &\vdots&\vdots&      &\vdots \\
0     &0     &\cdots &1     &0     &\cdots&0      \\
\end{array}\right),\qquad 
y \rightarrow \infty
\label{eq:bound1}
\end{eqnarray}
\begin{eqnarray}
 H^1(y)\rightarrow 
\sqrt{c}\left(
\begin{array}{ccccccc}
0     &\cdots&0     &1     &0&\cdots &0      \\
0     &\cdots&0     &0     &1&\cdots &0      \\
\vdots&      &\vdots&\vdots&\vdots &       &\vdots \\
0     &\cdots&0     &0     &0&\cdots &1      \\
\end{array}\right),\qquad 
y \rightarrow -\infty
\label{eq:bound2}
\end{eqnarray}

To solve the BPS equations, we define the following 
 $GL (N_{\rm C},{\bf C})$ 
group element $S(y)$ 
\begin{eqnarray}
 \Sigma +iW_y=S^{-1}(y)\partial _yS(y). 
\label{eq:slnc-ggroup}
\end{eqnarray}
The hypermultiplet scalar BPS equations (\ref{eq:BPSeq-H1}) 
and (\ref{eq:BPSeq-H2}) 
can be solved with this matrix function $S(y)$ as 
\begin{eqnarray}
H^1=S^{-1}(y)H_0 \; {e}^{My},\qquad H^2=0, 
\label{eq:hyper-sol-un}
\end{eqnarray}
with the constant matrix $H_0$ which is called the moduli 
matrix. 
We used the boundary conditions $H^2(y)\rightarrow 0$ 
at $y \rightarrow \pm\infty$ to fix $H^2(y)$ in the above 
solution \cite{INOS2}. 
Since the matrix function $S(y)$ in Eq.(\ref{eq:slnc-ggroup}) 
is defined up to a constant $GL (N_{\rm C},{\bf C})$ matrix $V$, 
equivalent descriptions for the 
hypermultiplet scalar $H^i$ and the vector multiplet scalar $\Sigma$ 
 are obtained by 
two sets of moduli matrices $H_0$ and $S$ that 
are related by $GL (N_{\rm C},{\bf C})$ 
transformations $V$ 
\begin{eqnarray}
(H_0, S) \sim (H'_0, S'), \qquad 
H_0\rightarrow H'_0=V H_0, \quad S\rightarrow S'=V S, 
\quad V\in GL (N_{\rm C},{\bf C}). 
\label{eq:world-v-sym-un}
\end{eqnarray}
We call this symmetry the $V$-equivalence relation. 
The space of the moduli matrix divided by 
 the $V$-equivalence relation is found \cite{INOS1} 
to be the complex 
Grassmann manifold $G_{N_{\rm F},N_{\rm C}}$ 
\begin{eqnarray}
\{H_0 | H_0 \sim V H_0, V \in GL(N_{\rm C},{\bf C})\} 
 \simeq G_{N_{\rm F},N_{\rm C}}
 \simeq {SU(N_{\rm F}) \over 
 SU(N_{\rm C}) \times SU(\tilde N_{\rm C}) \times U(1)}\,.
  \label{Gr}
\end{eqnarray}

In place of the gauge variant matrix function $S(y)$, 
we define a gauge invariant $N_{\rm C}\times N_{\rm C}$ matrix 
function 
\begin{eqnarray}
\Omega \equiv S S^\dagger. 
\label{eq:gauge-inv-un}
\end{eqnarray}
The remaining BPS equation (\ref{BPSeq-Sigma}) 
for the vector multiplet 
scalar $\Sigma$ can be rewritten in terms of $\Omega$ 
as {\it a master equation for the $U(N_{\rm C})$ gauge 
theory} \cite{INOS1}, \cite{INOS2} 
\begin{eqnarray}
 \partial _y\left(\Omega ^{-1}\partial _y\Omega \right)
=g^2c\left({\bf 1}_{\rm C}-\Omega ^{-1}\Omega _0\right),
\label{eq:master-eq-un}
\end{eqnarray}
where the source term is defined in terms of the moduli 
matrix $H_0$ as 
\begin{eqnarray}
\Omega _0\equiv c^{-1}H_0 \; {e}^{2My}H_0^\dagger . 
\label{eq:omega-0}
\end{eqnarray}

We need to specify the boundary conditions for $\Omega(y)$. 
For the maximal topological sector, the following quantity 
reduces to the unit matrix for the first 
$N_{\rm C}\times N_{\rm C}$ 
diagonal part in the limit 
of $y\rightarrow\infty$ 
\begin{eqnarray}
H^{1\dagger}(y)H^1(y)={e}^{My}H_0^{\dagger}\Omega(y)H_0 
\; {e}^{My}
\rightarrow 
{c}\left(
\begin{array}{ccccccc}
1     &0     &\cdots &0     &0     &\cdots&0      \\
0     &1     &\cdots &0     &0     &\cdots&0      \\
\vdots&      &       &\vdots&\vdots&      &\vdots \\
0     &0     &\cdots &1     &0     &\cdots&0      \\
0     &0     &\cdots &0     &0     &\cdots&0      \\
\vdots&\vdots&       &\vdots&\vdots&      &\vdots \\
0     &0     &\cdots &0     &0     &\cdots&0      \\
\end{array}\right),\; 
y \rightarrow \infty, 
\label{eq:bound3}
\end{eqnarray}
and to the unit matrix for the last $N_{\rm C}\times N_{\rm C}$ 
diagonal part in the limit 
of $y\rightarrow-\infty$ 
\begin{eqnarray}
H^{1\dagger}(y)H^1(y)={e}^{My}H_0^{\dagger}\Omega(y)H_0 
\; {e}^{My}
\rightarrow 
{c}\left(
\begin{array}{ccccccc}
0     &\cdots&0     &0     &0     &\cdots&0      \\
\vdots&      &\vdots&\vdots&\vdots&      &\vdots \\
0     &\cdots&0     &0     &0     &\cdots&0      \\
0     &\cdots&0     &1     &0     &\cdots&0      \\
0     &\cdots&0     &0     &1     &\cdots&0      \\
\vdots&      &\vdots&\vdots&\vdots&      &\vdots \\
0     &\cdots&0     &0     &0     &\cdots&1      \\
\end{array}\right),\; 
y \rightarrow -\infty .
\label{eq:bound4}
\end{eqnarray}

It has been conjectured \cite{INOS1} that there exists 
a unique solution of the master equation (\ref{eq:master-eq-un}) 
given the above boundary conditions (\ref{eq:bound3}) 
and (\ref{eq:bound4}). 
If this is proved, the solutions of the BPS equations 
(\ref{eq:BPSeq-H1})--(\ref{BPSeq-Y12}) in the 
$U(N_{\rm C})$ gauge theory with $N_{\rm F}$ 
flavors are completely characterized by 
the moduli 
matrices $H_0$ divided by the $V$-equivalence relation 
 (\ref{eq:world-v-sym-un}).

For the generic moduli matrix $H_0$, 
we cannot give a proof of the existence and uniqueness 
of the solution $\Omega(y)$ of the master equation 
(\ref{eq:master-eq-un}) at present. 
However, we can exploit our proof for the $U(1)$ 
gauge theory if the moduli matrix 
$H_0$ is of a restricted form, as we describe now. 
Let us note that the master equation (\ref{eq:master-eq-un}) 
transforms covariantly 
under the  world-volume transformation 
(\ref{eq:world-v-sym-un}), where the matrix 
$H_0 \; {e}^{2My}H_0{}^\dagger$ transforms with 
multiplication of 
constant matrices $V$ and  
$V^\dagger $ from both sides of this matrix. 
This $V$-equivalence relation $V$ allows us to 
diagonalize the matrix $H_0 \; {e}^{2My}H_0{}^\dagger$ 
at one point of the extra dimension, 
say, $y=y_0$. 
If the matrix $H_0 \; {e}^{2My}H_0{}^\dagger$ with this gauge 
fixing remains diagonal at every other points 
in the extra dimension $y\not=y_0$, we obtain 
\begin{eqnarray}
 H_0 \; {e}^{2My}H_0{}^\dagger =c\,{\rm diag.}
\left({\cal W}_1(y),{\cal W}_2(y),\cdots,
{\cal W}_{N_{\rm C}}(y)\right). 
\label{factorizable-case}
\end{eqnarray}
If this is valid, we call that moduli matrix $H_0$ 
as $U(1)$-factorizable \cite{INOS2}. 
Whether a given moduli matrix is $U(1)$ factorizable or not 
is an inherent characteristic of each moduli matrix $H_0$, 
and is independent of the choice of the 
initial coordinate $y_0$. 
Thus the $U(1)$-factorizability 
is a property attached to each point on the moduli space. 
If the moduli matrix is $U(1)$-factorizable, 
off-diagonal components of the 
matrix $H_0 \; {e}^{2My}H_0{}^\dagger $ vanishes 
at any point of the extra dimension $y$ by definition. 
Therefore each coefficient of 
${e}^{2m_Ay}$ in the off-diagonal components 
must vanish. 
Since we consider the case of non-degenerate masses, 
the condition for the 
$U(1)$-factorizability can be rewritten for 
each flavor $A$ as 
\begin{eqnarray}
 (H_0)^{r A}\left((H_0{})^{s A}\right)^*=0, 
\quad {\rm for~} r\not=s , 
\label{factorizable-moduli}
\end{eqnarray}
where the flavor index $A$ is not summed. 
Namely, $(H_0)^{rA}$ can be non-vanishing 
in only one color component $r$ 
for each flavor $A$.

To solve the master equation (\ref{eq:master-eq-un}) 
for the gauge invariant matrix function $\Omega$ 
in the non-Abelian $U(N_{\rm C})$ gauge theory 
with the $U(1)$-factorizable moduli, we are allowed 
to take an ansatz where only the 
diagonal components of the matrix $\Omega $ survive 
\begin{eqnarray}
 \Omega ={\rm diag.}\left({e}^{2 \psi _1},\,{e}^{2\psi _2},
\,\cdots,{e}^{2\psi _{N_{\rm C}}}\right),
\end{eqnarray} 
where $\psi_r(y)$'s are real functions. 
With this ansatz, the master equation (\ref{eq:master-eq-un}) 
for the $U(1)$-factorizable moduli with 
the condition (\ref{factorizable-case}) reduces to a set 
of the master equations 
for the Abelian gauge theory~\cite{INOS1} 
\begin{eqnarray}
 {d^2 \psi _r \over dy^2}
={g^2c\over 2}\left(1-{e}^{-2\psi _r}{\cal W}_r\right), 
\qquad 
{\rm for ~} r=1,2,\cdots N_{\rm C}, 
\label{U1BPS}
\end{eqnarray}
where the functions ${\cal W}_r(y)$ defined in 
(\ref{factorizable-case}) are given by 
\begin{eqnarray}
 {\cal W}_r=\sum_{A\in {\cal A}_r}{e}^{2 m_Ay} 
{|H_0^A|^2 \over \zeta}. 
\end{eqnarray}
${\cal A}_r$ is a set of flavors of the 
hypermultiplet scalars whose $r$-th 
color component is non-vanishing. 
Note that the condition (\ref{factorizable-moduli}) 
of the $U(1)$-factorizability can be rewritten as 
${\cal A}_r\cap{\cal A}_s=\emptyset $ for $r\not=s$. 
In this case, the vector multiplet scalars $\Sigma $ 
and the hypermultiplet scalars $H^{1rA}$ 
are given by~\cite{IOS1}  
\begin{eqnarray}
 \Sigma 
={\rm diag.}\left(\partial _y\psi _1,\,
\partial _y\psi _2,\,\cdots,\,
\partial _y\psi _{N_{\rm C}}\,\right) ,
\label{eq:factor-sigma}
\end{eqnarray}
\begin{eqnarray}
 H^{1rA} 
= {e}^{-\psi _r(y) +m_A y}H_0^A ,
\label{eq:factor-hyper}
\end{eqnarray}
with a gauge choice of $W_y=0$ and a phase choice of 
Im$ H^{1rA}=0$ at $y=\pm \infty$. 
Since the moduli parameters contained in the master equation 
(\ref{U1BPS}) for each $\psi _r$ are independent 
of each other, 
we find that our system of BPS equations for walls 
becomes a decoupled set of $N_{\rm C}$ systems of 
BPS equations of $U(1)$ gauge theories. 
This fact enables us to apply our method of proof 
for $U(1)$ gauge theories to this $U(1)$-factorizable 
case of the $U(N_{\rm C})$ gauge theory. 

Let us count the dimensions of the part of the moduli 
space with the $U(1)$-factorizable property. 
For each color component $r$, we denote the number of 
non-vanishing hypermultiplets in ${\cal A}_r$ as 
$f_r$. 
For the $r$-th color component, 
we obtain a $U(1)$ gauge theory with $f_r$ flavors. 
In the maximal topological sector, all the flavors should 
participate: $\sum_r f_r=N_{\rm F}$. 
Since every color component has to appear at 
vacua of both infinities, 
we obtain the complex dimension of the $U(1)$-factorizable 
part of the moduli space in the maximal topological sector 
to be 
\begin{eqnarray}
 {\rm dim}_{\bf C} {\cal M}_{U(1){\rm fact}} 
=\sum_{r=1}^{N_{\rm C}} (f_r-1)
= N_{\rm F}- N_{\rm C}. 
\label{eq:dim-fact}
\end{eqnarray}
We thus obtain a rather weak lower bound for the dimension
of the moduli space for the $U(N_{\rm C})$ gauge theory: 
\begin{eqnarray}
{\rm dim}_{\bf C} {\cal M}_{U(N_{\rm C})} \ge 
 N_{\rm F}- N_{\rm C}. 
\label{eq:dim-fact}
\end{eqnarray}

\section{
Proof of Existence and Uniqueness of Solution
}
\label{sc:existence}

In this section, we prove the existence and uniqueness of the 
BPS wall solutions to 
the $U(1)$ and $U(1)$-factorizable models. We first state 
our results in suitably renormalized parameters. 
We then carry out the proof using the method 
of calculus of variations and functional analysis.

For convenience, we relabel the variables and parameters 
of our master equations 
 (\ref{eq:master-u1}) and (\ref{U1BPS}) 
so that they are of the equivalent form\footnote{ 
For the $U(1)$ model, we denote $\lambda \equiv 2g^2\zeta$, 
$\omega_A\equiv 2m_A$, $u(y)\equiv -2\psi(y)$, 
$r_A\equiv \log(|H_0^A|^2/\zeta)$, and $M(y)\equiv 
e^{2W(y)}=\Omega_0(y)=\sum_A e^{2m_A y}|H_0^A|^2/\zeta$. 
}
\begin{equation}\label{eu1}
u''=\lambda(M(y) {e}^u-1),
\end{equation}
subject to the boundary conditions
\begin{equation}\label{eu2}
u(y)\to -\omega_1 y -r_1,\quad\quad y\to \infty,
\end{equation}
\begin{equation}\label{eu3}
u(y)\to -\omega_{N_{\rm F}} y-r_{N_{\rm F}},\quad\quad y\to -\infty,
\end{equation}
where $u'$ stands for $du/dy$ and
\begin{equation}\label{eu4}
M(y)=\sum_{A=1}^{N_{\rm F}} {e}^{\omega_A y +r_A},
\end{equation}
$\lambda>0, \omega_A, r_A$ are real constants so that $\omega_A$ 
($A=1,2,\cdots,{N_{\rm F}}$) satisfy the nondegeneracy condition 
\begin{equation}\label{eu5}
\omega_1>\omega_2>\cdots\omega_{N_{\rm F}}.
\end{equation}

For the equation (\ref{eu1}) subject to the boundary 
conditions (\ref{eu2}) and (\ref{eu3}), we have 

{\bf Theorem 4.1.} {\em 
The problem (\ref{eu1})--(\ref{eu3}) has a unique solution. 
Moreover, this solution also enjoys the estimates of 
the asymptotic exponential decay at $y\rightarrow \pm \infty$ 
}
\begin{eqnarray}
u(y)+(\omega_1 y+r_1)
&=&\mbox{O} ({e}^{-\lambda_1(1-\varepsilon)y})
\quad\mbox{\em as }
y\to\infty,\\
u(y)+(\omega_{N_{\rm F}} y +r_{N_{\rm F}})
&=&\mbox{O} ({e}^{\lambda_2 (1-\varepsilon)y})
\quad\mbox{\em as }
y\to-\infty,
\end{eqnarray}
{\em where $\varepsilon>0$ can be taken to be arbitrarily 
small and $\lambda_1$
and $\lambda_2$ are positive parameters defined by }\footnote{
The bound agrees with the previous result of the 
iterative approximation \cite{IOS1}. 
}
\begin{eqnarray}
\lambda_1
&=&\min\{\sqrt{\lambda},\omega_1-\omega_2\},\\
\lambda_2
&=&\min\{\sqrt{\lambda},
\omega_{{N_{\rm F}}-1}-\omega_{N_{\rm F}}\}.
\end{eqnarray}

We now proceed with the proof, adapting the main ideas 
from \cite{JT}.

Note that the conditions (\ref{eu2}), (\ref{eu3}), 
(\ref{eu5}) ensure that the right-hand side of (\ref{eu1}) 
vanishes at $y=\pm\infty$ because
\begin{eqnarray}
M(y)&=&{e}^{\omega_1y+r_1}(1+{e}^{(\omega_2-\omega_1)y+r_2-r_1}
+\cdots+{e}^{(\omega_{N_{\rm F}}-\omega_1)y
+r_{N_{\rm F}}-r_1}),\quad y>0,
\nonumber\\
M(y)&=&{e}^{\omega_{N_{\rm F}}y+r_{N_{\rm F}}}
(1+{e}^{(\omega_1-\omega_{N_{\rm F}})y+r_1-r_{N_{\rm F}}}
+\cdots+{e}^{(\omega_{{N_{\rm F}}-1}
-\omega_{N_{\rm F}})y+r_{{N_{\rm F}}-1}-r_{N_{\rm F}}}),\quad y<0.
\nonumber
\end{eqnarray}

To take into account of the boundary asymptotics, we 
introduce a translation, $u=u_0+v$,
where $u_0(y)$ has continuous second-order derivative and 
satisfies 
\begin{equation}\label{u0}
u_0(y)=-\omega_1 y -r_1\quad\mbox{if }y>y_0,
\quad u_0(y)=-\omega_{N_{\rm F}} y-r_{N_{\rm F}}
\quad\mbox{if } y<-y_0,
\end{equation}
where $y_0>0$ is a suitable constant to be determined shortly.
Then the equation (\ref{eu1}) becomes 
\begin{equation}\label{eu6}
v''=\lambda (Q(y){e}^v-1)+h(y),
\end{equation}
where $h(y)=-u_0''(y)$ is of compact support and 
$Q(y)={e}^{u_0(y)}M(y)$ has the representations 
\begin{eqnarray}
Q(y)&=&1+{e}^{(\omega_2-\omega_1)y+r_2-r_1}
+\cdots+{e}^{(\omega_{N_{\rm F}}-\omega_1)y
+r_{N_{\rm F}}-r_1},\quad y>y_0,
\label{Q+}\\
Q(y)&=&1+{e}^{(\omega_1-\omega_{N_{\rm F}})y+r_1-r_{N_{\rm F}}}
+\cdots+{e}^{(\omega_{{N_{\rm F}}-1}
-\omega_{N_{\rm F}})y+r_{{N_{\rm F}}-1}-r_{N_{\rm F}}},\quad y<-y_0.
\label{Q-}
\end{eqnarray}
Of course, $Q(y)>0$ everywhere, $Q(\pm\infty)=1$, and the 
boundary conditions (\ref{eu2}) and (\ref{eu3}) become the 
standard one,
\begin{equation}\label{eu7}
v=0\quad\mbox{at }\,\,\,y=\pm\infty.
\end{equation}

Since $u_0'(y)=-\omega_1$ for $y>y_0$ and 
$u_0'(y)=-\omega_{N_{\rm F}}$ for $y<-y_0$, it is clear that, 
when $y_0>0$ is sufficiently large, we can define 
$u_0'(y)$ for $-y_0\leq y\leq y_0$ to make 
$|u_0''(y)|$ as small as we please. 
In particular, we may achieve (say) 
\begin{equation}\label{eu0}
|h(y)|=|u_0''(y)|<\frac\lambda 2\quad\quad\mbox{for all } y.
\end{equation}
This assumption will be observed in the subsequent analysis.

It is clear that (\ref{eu6}) is the Euler--Lagrange equation 
of the action functional
\begin{equation}\label{eu8}
I(v)=\int \bigg\{
\frac12 (v')^2+\lambda Q(y)({e}^v-v-1)+\lambda (Q(y)-1) v+hv\bigg\},
\end{equation}
where and in the sequel, we use $\int$ to stand for the 
Lebesgue integral over the
whole real line $(-\infty,\infty)$ and we omit writing out 
the measure $dy$ when no risk of
confusion arises.

In order to accommodate the boundary condition (\ref{eu7}), 
we work on the standard Sobolev space
$W^{1,2}({\bf R})$= the completion of the set of all 
compactly supported real-valued
smooth functions over $\bf R$ under the norm
\[
\|f\|^2_{W^{1,2}({\bf R})}=\|f\|_2^2+\|f'\|_2^2,
\]
where we use $\|\cdot\|_p$ ($p\geq1$) to denote the integral norm
\[
\|f\|_p=\bigg(\int |f(y)|^p \bigg)^{\frac1p}.
\]
Use $C({\bf R})$ to denote the space of continuous 
functions over $\bf R$ vanishing at $\pm\infty$, equipped 
with the standard pointwise norm
\[
\|f\|_{C({\bf R})}=\sup_{-\infty<y<\infty}|f(y)|.
\]
Then an immediate application of the Schwartz inequality 
yields the continuous embedding 
$W^{1,2}({\bf R})\subset C({\bf R})$ with
\begin{equation}\label{eu9}
\|f\|_{C(\bf R)}\leq \|f\|_{W^{1,2}({\bf R})},
\quad f\in W^{1,2}({\bf R}).
\end{equation}

Using (\ref{eu9}), we see that the functional (\ref{eu8}) 
is a well-defined, continuously differentiable, functional 
over $W^{1,2}({\bf R})$, which is also strictly convex.

In the following, we show that (\ref{eu8}) has a critical 
point in $W^{1,2}({\bf R})$ by minimizing (\ref{eu8}) 
over $W^{1,2}({\bf R})$.

Recall the following one-dimensional Gagliardo--Nirenberg 
inequality:
\begin{equation}\label{eu10}
\int |f|^{p+1}\leq C(p)\bigg(\int f^2
\bigg)^{\frac{p+3}4}\bigg(\int (f')^2\bigg)^{\frac{p-1}4},
\quad
\quad f\in W^{1,2}({\bf R}),
\end{equation}
where $p>1$ and $C(p)$ is a positive constant depending 
only on $p$.

Note that, in view of the Schwartz inequality and 
Gagliardo--Nirenberg inequality (\ref{eu10}) (with $p=3$),
we have
\begin{eqnarray}
&&\bigg(\int v^2 \bigg)^2\nonumber\\
&&=\bigg(\int
\frac{|v|}{1+|v|}(1+|v|)|v| \bigg)^2 \nonumber\\
&&\leq\int \frac{v^2}{(1+|v|)^2} \int (1+|v|)^2 v^2 
\nonumber\\
&&\leq 2\int \frac{v^2}{(1+|v|)^2} \int (v^2+v^4) 
\nonumber\\
&&\leq C_1\int \frac{v^2}{(1+|v|)^2}
\bigg(\int v^2 +\bigg(\int v^2 \bigg)^{\frac32}
\bigg(\int (v')^2\bigg)^{\frac12}\bigg) \nonumber\\
&&\leq\frac12\bigg(\int v^2 \bigg)^2
+C_2\bigg(\int\frac{v^2}{(1+|v|)^2}\bigg)^2
+ C_3\bigg(\bigg(\int\frac{v^2}{(1+|v|)^2}\bigg)^6
+\bigg(\int (v')^2\bigg)^6\bigg).\nonumber
\end{eqnarray}
Consequently, we have
\begin{equation}{\label{eu13}}
\|v\|_2\leq C_0 \bigg(1+\int\frac{v^2}{1+|v|}
+\int (v')^2\, dy\bigg)^{\frac32},
\end{equation}
where $C_0>0$ is a suitable constant.

Define $J(v)=(DI(v))(v)$ 
(the Fr\'{e}chet derivative) for $v\in W^{1,2}({\bf R})$. 
Then 
\begin{eqnarray}
J(v)&=&
\lim_{t\to 0}\frac{I({v}+t{v})-I({v})}t
=(DI({v}))({v}) \nonumber \\
&=&\int \bigg\{ (v')^2 +\lambda Q(y)({e}^v-1)v+q(y)v\bigg\},
\label{eu11}
\end{eqnarray}
where $q(y)=\lambda (Q(y)-1)+h(y)$ vanishes at 
$y=\pm\infty$ exponentially fast.

Let $v^+$ and $v^-$ be the positive and negative parts of 
$v$ respectively. That is, $v=v^+-v^-$ and
\[
v^+=\max\{v,0\}=\frac12(v+|v|),
\quad v^-=\max\{0,-v\}=\frac12(|v|-v).
\]
Then the functional $J(v)$ defined in (\ref{eu11}) may be 
rewritten as
\begin{equation}\label{eu14}
J(v)=\|v'\|_2^2+J_1(v)+J_2(v),
\end{equation}
where
\begin{eqnarray}
J_1(v)&=&\int\{\lambda Q(y)({e}^{v^+}-1)v^++q(y) v^+\},
\nonumber\\
J_2(v)&=&\int\{\lambda Q(y)({e}^{-v^-}-1)(-v^-)-q(y) v^-\}.
\nonumber
\end{eqnarray}
Recall that $Q(y)\geq1, ({e}^{v^+}-1)v^+\geq (v^+)^2, 
({e}^{-v^-}-1)(-v^-)\geq
(v^-)^2/(1+|v^-|)$. Hence
\begin{eqnarray}
J_1(v)&\geq&\lambda\|v^+\|^2_2-\|q\|_2\|v^+\|_2\nonumber\\
&\geq&\frac\lambda2\|v^+\|^2_2-\frac1{2\lambda}\|q\|^2_2,
\label{eu15}\\
J_2(v)&\geq&\int\lambda Q(y)\frac{(v^-)^2}{1+|v^-|}
-\int q(y)\frac{v^-}{1+|v^-|}(1+|v^-|)\nonumber\\
&\geq&\int \lambda Q(y)\frac{(v^-)^2}{1+|v^-|}
-\int |q(y)| -\int (\lambda(Q(y)-1)+h(y))
\frac{(v^-)^2}{1+|v^-|}\nonumber\\
&\geq&\int (\lambda-|h(y)|)\frac{(v^-)^2}{1+|v^-|}
-\int |q(y)| \nonumber\\
&\geq&\frac\lambda2\int\frac{(v^-)^2}{1+|v^-|}
-\int |q(y)|,\label{eu16}
\end{eqnarray}
where we have used (\ref{eu0}).
Inserting (\ref{eu15}) and (\ref{eu16}) into (\ref{eu14}) 
and applying (\ref{eu13}), we arrive at
\begin{eqnarray}\label{eu17}
J(v)&\geq& \|v'\|^2_2+\frac\lambda2\int\frac{v^2}{1+|v|}
-C_4,\nonumber\\
&\geq&\frac12\|v'\|^2_2
+\min\bigg\{\frac12,\frac\lambda2\bigg\}\bigg(\|v'\|^2_2
+\int\frac{v^2}{1+|v|}
\bigg)-C_4\nonumber\\
&\geq&\frac12\|v'\|^2_2
+\min\bigg\{\frac12,\frac\lambda2\bigg\} 
C^{-2/3}_0\|v\|_2^{2/3}-C_5,
\end{eqnarray}
where the constants $C_4, C_5>0$ depend only on $\lambda$ 
and $q(y)$. Note also that 
$\|v'\|^{2/3}_2\leq(1/3)\|v'\|_2^2+(2/3)$. 
We can rewrite (\ref{eu17}) more evenly as
\begin{eqnarray}\label{eu18}
J(v)&\geq&C_6(\|v\|^{2/3}_2+\|v'\|^{2/3}_2)-C_7\nonumber\\
&\geq& C_8\|v\|^{2/3}_{W^{1,2}({\bf R})}-C_9,
\end{eqnarray}
where $C_6, C_7, C_8, C_9$ are some positive constants.

With the above estimates, we are now ready to do 
minimization following a standard path.

In view of (\ref{eu18}), let $R>0$ be such that
\begin{equation}\label{eu19}
\inf\{ J(v)\,|\, v\in W^{1,2}({\bf R}),
\,\|v\|_{W^{1,2}({\bf R})}=R\}\geq1,
\end{equation}
and consider the minimization problem
\begin{equation}\label{eu20}
\sigma=\inf \{ I(v)\,|\,\|v\|_{W^{1,2}({\bf R})}\leq R\}.
\end{equation}

Let $\{v_n\}$ be a minimizing sequence of the problem 
(\ref{eu20}). Since $\{v_n\}$ is bounded in 
$W^{1,2}({\bf R})$, by extracting a subsequence if 
necessary, we may assume that $\{v_n\}$ is also weakly 
convergent. 
Let $\tilde{v}$ be its weak limit in $W^{1,2}({\bf R})$. 
Since $I(\cdot)$ is a continuously differentiable 
functional over $W^{1,2}({\bf R})$ and convex, 
$I(\cdot)$ is weakly lower semicontinuous. 
Hence $I(\tilde{v})\leq \lim_{n\to\infty} I(v_n)=\sigma$. 
Of course, $\|\tilde{v}\|_{W^{1,2}({\bf R})}\leq R$ 
because the norm of $W^{1,2}({\bf R})$ is also weakly 
lower semicontinuous. Hence $\tilde{v}$ solves 
(\ref{eu20}). 
That is, $I(\tilde{v})=\sigma$.
To show that $\tilde{v}$ is a critical point of 
$I(\cdot)$, we need to show that $\tilde{v}$ is interior, 
or $\|\tilde{v}\|_{W^{1,2}({\bf R})}<R$.

Otherwise, if $\|\tilde{v}\|_{W^{1,2}({\bf R})}=R$, 
then, by (\ref{eu19}), we have 
\begin{eqnarray}\label{eu21}
\lim_{t\to 0}\frac{I(\tilde{v}-t\tilde{v})-I(\tilde{v})}t
&=&-(DI(\tilde{v}))(\tilde{v})\nonumber\\
&=&-J(\tilde{v})\leq-1.\nonumber
\end{eqnarray}
In particular, when $t>0$ is sufficiently small, we have 
$I(\tilde{v}-t\tilde{v})<I(\tilde{v})=\sigma$. 
On the other hand, 
$\|\tilde{v}-t\tilde{v}\|_{W^{1,2}({\bf R})}=(1-t)R<R$. 
These two facts violate the definition of $\sigma$ made 
in (\ref{eu20}).

Therefore $\tilde{v}$ is interior. 
Consequently, it is a critical point of $I(\cdot)$ in 
$W^{1,2}({\bf R})$, which is a weak solution of (\ref{eu6}). 
The standard elliptic regularity theory shows that this 
gives rise to a $C^\infty$-solution of the original 
equation (\ref{eu1}) subject to the boundary conditions
(\ref{eu2}) and (\ref{eu3}).

The strict convexity of $I(\cdot)$ already implies that 
$I(\cdot)$ can have at most one critical point in 
$W^{1,2}({\bf R})$. 
Hence $I(\cdot)$ has exactly one critical point in 
$W^{1,2}({\bf R})$ and the uniqueness of a solution to 
(\ref{eu1})--(\ref{eu3}) or (\ref{eu6}) and (\ref{eu7}) 
follows. 
In fact, such a uniqueness result follows from the 
structure of the equation in a more straightforward way: 
if $v_1$ and $v_2$ are two solutions of (\ref{eu6}) and 
(\ref{eu7}), then $w=v_1-v_2$ satisfies 
$w''=\lambda Q(y) {e}^{\xi(y)} w$ where $\xi(y)$ lies 
between $v_1(y)$ and $v_2(y)$. Since $w=0$ at 
$y=\pm\infty$ and $Q(y){e}^{\xi(y)}>0$ for any $y$, 
we must have $w(y)\equiv0$.

Finally, we estimate the asymptotic exponential decay rates 
of the solution of (\ref{eu6}) and 
(\ref{eu7}) near $y=\pm\infty$. For $y>y_0$, we see that 
(\ref{eu6}) takes the form
\begin{equation}
v''=\lambda Q(y)({e}^v-1)+\lambda (Q(y)-1).\quad \label{eu23}
\end{equation}
Introduce a comparison function 
\begin{equation}\label{eu24}
V=C {e}^{-\sqrt{\omega}(1-\varepsilon)y},\quad \omega=
\min\{\lambda,(\omega_1-\omega_2)^2\},\quad 0<\varepsilon<1.
\end{equation}
Then $V$ satisfies
$V''=\omega (1-\varepsilon)^2 V$.
In view of this, (\ref{eu23}), and (\ref{eu24}), we have
\begin{equation}\label{eu25}
(v\pm V)''=\lambda Q(y)e^{\xi(y)}(v\pm V)
+\lambda(Q(y)-1)\mp(\lambda Q(y)e^{\xi(y)}-
\omega(1-\varepsilon)^2) C e^{-\sqrt{\omega}(1-\varepsilon)y},
\end{equation}
where $\xi(y)$ lies between $0$ and $v(y)$. 
Assume that the constant $C$ in (\ref{eu24}) satisfies 
$C\geq1$ (say). 
Since $v(y)\to0$ and $Q(y)-1=\mbox{O}
(e^{-(\omega_1-\omega_2)y})$ as $y\to\infty$, 
we can find a sufficiently large $y_1>0$
so that
\begin{equation}\label{eu26}
\lambda(Q(y)-1)-(\lambda Q(y)e^{\xi(y)}-
\omega(1-\varepsilon)^2) C 
e^{-\sqrt{\omega}(1-\varepsilon)y}<0,\quad y>y_1.
\end{equation}
Combining (\ref{eu25}) and (\ref{eu26}), we arrive at
\begin{equation}\label{eu27}
(v+V)''<\lambda Q(y)e^{\xi(y)}(v+V),\quad y>y_1.
\end{equation}
Of course, we may choose the constant $C\geq1$ in the 
definition of the function $V$ (see (\ref{eu24})) 
sufficiently large so that $(v+V)(y_1)\geq0$. 
Using this condition, the fact that $v+V=0$ at 
$y=\infty$, and (\ref{eu27}), we get $(v+V)(y)>0$ 
for all $y>y_1$.

On the other hand, since $Q(y)>1$ for $y>y_0$ 
(see (\ref{Q+})), in view of (\ref{eu24}) and (\ref{eu25}) 
again, we see that there is a sufficiently large $y_2>y_0$
so that
\begin{equation}\label{eu28}
(v-V)''>\lambda Q(y)e^{\xi(y)}(v-V),\quad y>y_2.
\end{equation}
Again, we may choose the constant $C\geq1$ (say) 
in the definition of the function $V$ (see (\ref{eu24})) 
sufficiently large so that $(v-V)(y_2)\leq0$. 
Using this condition, the fact that $v-V=0$ at 
$y=\infty$, and (\ref{eu28}), we get $(v-V)(y)<0$ 
for all $y>y_2$.

In summary, we have obtained the expected asymptotic 
exponential decay estimate 
$|v(y)|<V(y)=Ce^{-\sqrt{\omega}(1-\varepsilon)y}$ for 
$y\to\infty$.

A similar argument leads to the exponential 
decay estimate for $v(y)$ as $y\to-\infty$.

The proof of Theorem 4.1 is now complete. 

\subsection*{The Acknowledgments}

N.S. wishes to thank a fruitful collaboration 
with Minoru Eto, Youichi Isozumi, Muneto Nitta, 
Keisuke Ohashi, Kazutoshi Ohta, Yuji Tachikawa, and 
David Tong. 
He is also benefitted from useful 
communications with Jarah Evslin and David Tong 
on the solvability of the master equations. 
N.S. is supported in part by Grant-in-Aid for Scientific 
Research from the Ministry of Education, Culture, Sports, 
Science and Technology, Japan No.16028203 for 
the priority area ``origin of mass'' and No.17540237. 
Y.Y.~was supported in part by NSF grant DMS--0406446.

\newcommand{\J}[4]{{\sl #1} {\bf #2} (#3) #4}
\newcommand{\andJ}[3]{{\bf #1} (#2) #3}
\newcommand{\AP}{Ann.\ Phys.\ (N.Y.)}
\newcommand{\MPL}{Mod.\ Phys.\ Lett.}
\newcommand{\NP}{Nucl.\ Phys.}
\newcommand{\PL}{Phys.\ Lett.}
\newcommand{\PR}{ Phys.\ Rev.}
\newcommand{\PRL}{Phys.\ Rev.\ Lett.}
\newcommand{\PTP}{Prog.\ Theor.\ Phys.}
\newcommand{\hep}[1]{{\tt hep-th/{#1}}}

\end{document}